\documentclass[onecolumn, draftcls, 12pt]{IEEEtran}

\usepackage{amsmath}
\usepackage{cite}
\usepackage{amssymb}
\usepackage{amsthm}
\usepackage{epsfig}
\usepackage{psfrag}
\usepackage{graphicx}
\usepackage{mathrsfs}
\usepackage{color}
\usepackage{epsfig}
\usepackage{epstopdf}
\usepackage{slashbox,booktabs,amsmath}
\usepackage{subfigure}
\usepackage{float}	
\usepackage[center]{caption}

\input{epsf}

\begin{document}
\title{From DC-Biased to DC-Informative Optical OFDM}

\author{Qian Gao,~Chen Gong,~Rui Wang, Zhengyuan Xu and Yingbo Hua*
\IEEEcompsocitemizethanks{\IEEEcompsocthanksitem *Q.~Gao, C.~ Gong and Z.~Xu are with University of Science and Technology of China, Hefei, 230022, China. R.~Wang is with Tongji University, Shanghai, 200092, China. Y.~Hua is with University of California, Riverside, CA, 92507, USA.}}

\maketitle

\IEEEpeerreviewmaketitle

\begin{abstract}
We propose a novel modulation scheme for intensity modulation and direct detection (IM/DD) based optical communication system employing orthogonal frequency division multiplexing (OFDM). This method utilizes the DC-bias, which typically is discarded at the receiver-end, to carry information to achieve higher power efficiency. By formulating and solving a convex optimization problem, a constellation in high dimensional space is designed offline for the input of the transmitter-side inverse fast Fourier transform (IFFT) block. We point out that one can choose partial or full DC power for information transmission. Under the condition that the spectrum efficiency is fixed and attainable, this method bears notable power gain over traditional DC-biased optical OFDM (DCO-OFDM).
\end{abstract}

\begin{IEEEkeywords}
Optical communication, OFDM, Informative DC, IM/DD, constellation design.
\end{IEEEkeywords}

\section{Introduction}
Optical communication using visible light, infrared, or ultraviolet are promising candidates to provide high-speed indoor/outdoor wireless access
on non-regulated frequency bands \cite{IEEE,Xu}. Unlike the radio frequency communication (RFC) systems, the light intensity is controlled to convey information in optical communications.
The baseband signals are required to be positive and real,
which impose some constraints on applying existing radio-frequency modulation schemes,
such as the orthogonal frequency division multiplexing (OFDM).
Among plenty of schemes proposed for Optical OFDM (O-OFDM), the asymmetrically-clipped optical OFDM (ACO-OFDM) and DC-biased optical OFDM (DCO-OFDM) proposed by J. Armstrong etc. are the most popular \cite{5Jean}. However, the ACO-OFDM scheme suffers from reduced spectral efficiency since only 1/4 of the subcarriers carry information, and the DCO-OFDM scheme suffers from low power efficiency due to use of a large DC-bias to compensate for the negative peak.

Significant efforts have been made to improve the ACO-OFDM and DCO-OFDM, such as bit-loading, adaptive modulation methods, etc~\cite{Jean,Shlomi,Kimura}.
However, one common issue is the employment of a non-informative DC-bias (for DCO-OFDM), which is subtracted before demodulation at the receiver-side.
It reduces the system power efficiency significantly.
Therefore, we propose to use the DC-bias as one dimension of information basis, and design a joint constellation across multiple subcarriers before the IFFT.
Such scheme is termed DC-Informative Optical OFDM (DCIO-OFDM).
Note that our constellation points are sphere-packed in high dimensional space, which shows a more compact structure than lower dimensional counterparts, e.g. when independent $M_i-QAM$ ($M_i$-ary quadrature amplitude modulation) is applied for each subcarrier.

It is worth noting that either full or partial DC-bias can be utilized as information basis with DCIO-OFDM, termed full DCIO-OFDM and partial DCIO-OFDM, respectively.
The system power gain is expected to grow with the ratio of the informative DC power over the total DC power.
Specifically, adopting partial DCIO-OFDM indicates a hybrid scheme between DCIO-OFDM and DCO-OFDM.
The system designers can choose an optimal informative DC power ratio according to the desired system configurations.

The remainder of this letter is organized as follows.
In Section~II, we show the system diagram of the traditional DCO-OFDM and the proposed DCIO-OFDM.
In Section III, we formulate a convex optimization problem to optimize the constellation for DCIO-OFDM that minimizes system error rate.
Numerical results are provided in Section~IV.
Finally, Section V concludes this paper.

\vspace{.1in}

\begin{figure*}[htbp!]
\centering
\includegraphics[width=15cm]{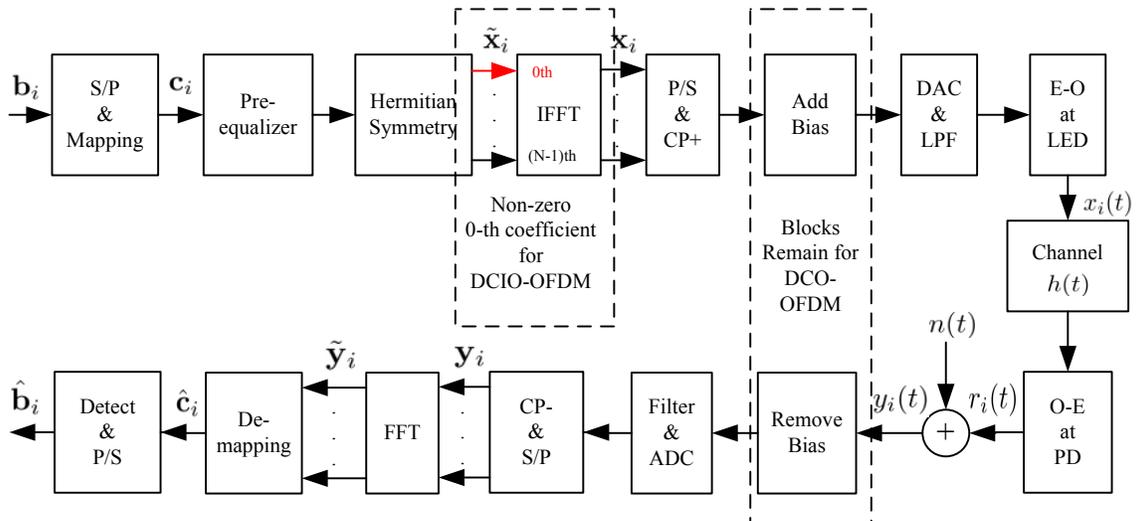}
\parbox{15cm}{\caption{Block diagram of the DCO/DCIO-OFDM scheme.}}
\end{figure*}

\section{From DCO-OFDM to DCIO-OFDM}
\subsection{Signal Model}
Consider the block diagrams of IM/DD based DCO-OFDM and DCIO-OFDM schemes, as shown in Fig. 1. The transmitter (Tx) can be light-emitting diodes (LED) or lasers within a broad optical spectrum depending on application scenario.
The frequency-domain discrete channel model that links input symbols $\tilde{\mathbf{x}}$ to transmitter IFFT  to output symbols $\tilde{\mathbf{y}}$ of receiver FFT is given as\footnote{The linearity between IFFT at transmitter and FFT at receiver is not disturbed as long as the convolution of the time-domain channel and a non-informative DC-bias (defined later) is removed before prefixing is abandoned.}
\begin{align}
\tilde{\mathbf{y}}&=\gamma\eta\mathbf{W}\mathbf{H}\mathbf{W}^H\tilde{\mathbf{x}}+\mathbf{W}\mathbf{n}\notag\\
&=\gamma\eta\mathbf{W}\mathbf{W}^H\boldsymbol{\Sigma_H}\mathbf{W}\mathbf{W}^H\tilde{\mathbf{x}}+\tilde{\mathbf{n}}\notag\\
&=\gamma\eta\boldsymbol{\Sigma_H}\tilde{\mathbf{x}}+\tilde{\mathbf{n}},
\end{align}
where $\tilde{\mathbf{x}}$ is the transmitter-side frequency-domain (independent) complex symbols of size $N\times 1$ where $N$ is the length of IFFT, $\gamma$ is the electrical to optical and $\eta$ is the optical to electrical conversion factors respectively\footnote{We assume $\gamma\eta=1$ for simplicity in this paper without loss of generality.}, $\boldsymbol{\Sigma}_{H}$ is an $N\times N$ diagonal matrix obtained from a decomposition of the circular channel $\mathbf{H}$, $\tilde{\mathbf{y}}$ is the reconstructed frequency-domain symbols after receiver-side FFT, $\mathbf{n}$ is an additive white Gaussian noise (AWGN) representing the sum of thermal and shot noises. The matrix $\mathbf{W}$ is an $N\times N$ DFT matrix with elements $w_{n,k}=\frac{1}{\sqrt{N}}e^{\frac{2\pi j\cdot nk}{N}}$, which is unitary and thus multiplying $\mathbf{W}$ with $\mathbf{n}$ does not change the statistics of $\mathbf{n}$.

Note that for an \textit{N}-point DCO-OFDM scheme \cite[Eq.1]{5Jean}, the frequency domain (FD) symbols have the following structure
\begin{equation}
\tilde{\mathbf{x}}=[0~ \tilde{x}_1~ \ldots~ \tilde{x}_{N/2-1}~ 0~ \tilde{x}_{N/2-1}^*~ \ldots \tilde{x}_1^*]^T,
\end{equation}
where $x_0$ and $x_{N/2}$ are set as zeros, and $x_{i\in[1,N/2-1]}=x_{N-i}^*$ is assumed to guarantee the realness of the bipolar output of IFFT. To make $\mathbf{x}$ unipolar, a DC-bias vector is added as follows
\begin{equation}
\hat{\mathbf{x}}=\mathbf{x}+B_{DC,z\text{dB}}\cdot\mathbf{1},
\end{equation}
where $\mathbf{1}$ is an $N\times 1$ all one vector, $B_{DC,z\text{dB}}=k\sqrt{\mathbb{E}\{x(t)^2\}}$ is defined as a bias of $z=10\log_{10}(k^2+1)$dB \cite{5Jean}, where $k$ is a proportionally constant and $\mathbb{E}\{x(t)^2\}$ is the expectation of the power of time-domain signal after digital to analog (DA) conversion. Here for simplicity we assume the DA is done using a zero-order hold circuit, such that the negative peak values of $\mathbf{x}$ and $x(t)$ are the same.

Assume that the constellation is jointly designed across only part of the sub-carrier.
Let $N_J$ be the number of independent sub-carriers other than the $0^{th}$ subcarrier, that are jointly designed with the $0^{th}$ subcarrier.
The value of $\tilde{x}_{N/2}$ is set to be zero.
Note that, due to the conjugate symmetry of the DCO-OFDM symbols before the IFFT,
the number of subcarriers involved into the joint design is $2N_J + 1$.
Let $M$ be the number of joint constellation points.
Note that, for a full DCIO-OFDM, all independent subcarriers are jointly designed, i.e., $N_J = N/2-1$;
and for a partial DCIO-OFDM, some independent subcarriers are designed jointly and others are designed independently,
i.e., $N_J < N/2-1$.
For a full DCIO-OFDM, there is no ``bias adding'' operation, since with properly designed constellation symbols,
the output of IFFT $\mathbf{x}_i$ is ensured to be non-negative.

Additional features of the proposed scheme include:
\begin{itemize}
\item Mapping/De-mapping are done based on the Binary Switching Algorithm (BSA) that minimizes the number of bits in error with one symbol error \cite{Schreckenbach};
\item No signal clipping is needed as long as (convex) dynamic range constraints are imposed on $\mathbf{x}_i$ with the design.
\end{itemize}

\subsection{Constellation Points Design}
Note that, for DCO-OFDM, an equivalent scheme is feasible by setting $\tilde{x}_0=\sqrt{N}B_{DC,z\text{dB}}$ and adding no DC after DA,
as long as $B_{DC,z\text{dB}}$ is estimated before IFFT instead of being measured after DA.
This motivates us to wonder how the $0$-th subcarrier can be better utilized, e.g. to be made adaptive to carry information bits.

We propose a joint mapping scheme instead, and the combined information bit sequences of length $N_b$ are jointly mapped to a \textit{real} constellation matrix
\begin{eqnarray}
\mathcal{C}&=&
\begin{bmatrix}
c^{(1)}_{1} & c_{1}^{(2)} & \ldots & c_{1}^{(M)}\\
c^{(1)}_{2} & c_{2}^{(2)} & \ldots & c_{2}^{(M)}\\
\vdots & \vdots & \ddots & \vdots\\
c_{2N_J+1}^{(1)} & c_{2N_J+1}^{(2)} & \ldots & c_{2N_J+1}^{(M)}\\
\end{bmatrix} \nonumber \\
&\triangleq& [\mathbf{c}^{(1)}, \mathbf{c}^{(2)}, ..., \mathbf{c}^{(M)}],
\end{eqnarray}
where the $m$-th ($m\in[1,M]$) column vector $\mathbf{c}^{(M)}$ is a constellation point in a $2N_J+1$ dimensional space, $M=2^{N_b}$ is the constellation size,
and $N_J\leq N/2 - 1$ is the number of subcarriers allocated for the joint constellation design part.
If $N_J=N/2-1$, we term the associated scheme as full DCIO-OFDM; and if $N_J<(N-1)/2$, it is termed partial DCIO-OFDM,
and the rest $N-N_J$ subcarriers are still independently applying, e.g. $M_i-QAM$. The constellation signal set needs to be properly designed such that $\mathbf{x}$ is non-negative guaranteed.

Note that $c_{1}^{(m)}=\tilde{x}^{(m)}_0$ are just the real adaptive DC-bias mentioned. Subcarrier $k\in[1,N_J]$ takes the value from
\begin{equation}
\tilde{x}_{k}^{(m)}=c_{2k}^{(m)}+jc_{2k+1}^{(m)};
\end{equation}
and to guarantee realness after IFFT, subcarrier $N-k$ takes value from
\begin{equation}
\tilde{x}_{N-k}^{(m)}=c_{2k}^{(m)}-jc_{2k+1}^{(m)}.
\end{equation}

For the $n$-th IFFT output, the joint constellation component can be derived based on the
DC and the $N_J$ subcarriers applying the $m$-th joint constellation symbol,
which is constrained to be nonnegative, i.e.,
\begin{align}
x^{(m)}_{n,Joint}&=\sum_{k=0}^{N_J}w_{n,k}\tilde{x}_k^{(m)}+\sum_{k=1}^{N_J}w_{n,k}^*
\tilde{x}_{N-k}^{(m)}\notag\\
&=\boldsymbol{\phi}_n^T\mathbf{c}^{(m)}\geq 0\qquad \forall n\in[1,N],\label{8}
\end{align}
where $\boldsymbol{\phi}_n$ is given by
\begin{equation*}
\boldsymbol{\phi}_n=2\cdot\bigg[\frac{1}{2},\Re(w_{n,1}),-\Im(w_{n,1})
\ldots \Re(w_{n,N_J}),-\Im(w_{n,N_J})\bigg]^T
\end{equation*}
and $\Re(a)$ and $\Im(a)$ denote the real and imaginary part of $a$, respectively.
Denote the $n$-th IFFT output corresponding to independent constellation design as $x_{n,Indep}$.
Thus, the $n$-th output of IFFT is the sum over the joint design component and the independent design components, given as follows,
\begin{equation}
x_{n,Comb}^{(m)}=x^{(m)}_{n,Joint}+x_{n,Indep}.
\end{equation}

Since $x_{n,Comb}^{(m)}$ is still bipolar due to the independent modulation component $x_{n,Indep}$,
an additional bias should be added after the DA. The equivalent expression in discrete domain is
\begin{equation}
x_n^{(m)}=x^{(m)}_{n,Joint}+x_{n,Indep}+B^\prime_{DC}\qquad \forall n\in[1,N],
\end{equation}
where the DC value $B^\prime_{DC}<B_{DC,z\text{dB}}$ is expected due to the
joint design components $x^{(m)}_{n,Joint}$.

\section{Problem Formulation}
We propose the optimal design of a stacked constellation vector of size $M(2N_J+1)\times 1$, denoted as
\begin{equation}
\mathbf{c}^{(s)}=[\mathbf{c}^{(1)T},\mathbf{c}^{(2)T}\ldots \mathbf{c}^{(M)T}]^T,
\end{equation}
by maximizing the minimum Euclidean distance (MED) among $\mathbf{c}^{(m)}$, subject to a non-negative IFFT output constraint
and a total energy constraint.

\subsection{The objective function}
To guarantee that the MED is larger than or equal to $d_{min}$, the following constraint must be satisfied,
\begin{equation}
\mathbf{c}^{(s)T}\mathbf{E}_{pq}\mathbf{c}^{(s)}\geq d_{min}^2,~~p,q\in[1,\frac{M(M-1)}{2}],~p<q,
\end{equation}
where $\mathbf{E}_{pq}$ is given by \cite{Beko12}
\begin{equation}
\mathbf{E}_{pq}=\mathbf{E}_p^T\mathbf{E}_p-\mathbf{E}_p^T\mathbf{E}_q-
\mathbf{E}_q^T\mathbf{E}_p+\mathbf{E}_q^T\mathbf{E}_q,\label{12}
\end{equation}
 where $\otimes$ denotes the Kronecker product, $\mathbf{I}_{M}$ is an $M\times M$ identity matrix, and $\mathbf{e}_p=[0,\ldots,1,\ldots,0]^T$ with a $1$ as the $p$-th element. Based on a linear approximation at a point $\mathbf{c}^{(s)T}_0$, the non-convex constraints in \eqref{12} can be turned into convex ones, i.e.
\begin{eqnarray} \label{equ.Approximation}
\mathbf{c}^{(s)T}\mathbf{E}_{pq}\mathbf{c}^{(s)} &\cong& 2\mathbf{c}^{(s)T}_0\mathbf{E}_{pq}\mathbf{c}^{(s)}-\mathbf{c}^{(s)T}_0\mathbf{E}_{pq}\mathbf{c}^{(s)}_0 \nonumber \\
&\geq& d^2_{min}, \forall (p,q).
\end{eqnarray}

\subsection{The Non-negative Constraint}
As in \eqref{8}, the IFFT of the joint constellation part needs to be non-negative, i.e.,
\begin{equation}
\boldsymbol{\phi}_n^T\mathbf{J}^{(m)}\mathbf{c}^{(s)}\geq 0, \qquad\forall (n,m),
\end{equation}
where $\mathbf{J}^{(m)}$ is a selection matrix such that $\mathbf{c}^{(m)}=\mathbf{J}^{(m)}\mathbf{c}^{(s)}$.
Clearly, it is a convex constraint in $\mathbf{c}^{(s)}$.

\subsection{The Average Electrical Power Constraint}
We constrain that the average electrical power is $P_{a}$.
Let $P^{(m)}$ be the power of the $m$-th joint modulated subcarriers including the conjugated subcarriers.
Assume the same power $P_{Indep}$ for all $N - 2N_J - 2$ independent subcarriers.
Then, the average electrical power of OFDM blocks is calculated as follows,
\begin{eqnarray}
P_a&=&\frac{1}{M}\sum_{m=1}^MP^{(m)}+(N - 2N_J - 2)P_{Indep}\nonumber\\
&=&\frac{1}{M}\bigg[\mathbf{c}^{(s)T}\mathbf{c}^{(s)}+
\sum_{m=1}^M\mathbf{c}^{(s)T}\mathbf{K}^{(m)T}\mathbf{K}^{(m)}\mathbf{c}^{(s)}\bigg] \nonumber\\
&& \ \ +M_fP_{Indep}\nonumber\\
&\triangleq& \mathbf{c}^{(s)}\bar{\mathbf{K}}\mathbf{c}^{(s)}+M_fP_{Indep},
\end{eqnarray}
where $\frac{1}{M}\mathbf{c}^{(s)T}\mathbf{c}^{(s)}$ is the average sum power of the adaptive DC and from subcarrier $1$ to $N_J$; $\frac{1}{M}\sum_{m=1}^M\mathbf{c}^{(s)T}\mathbf{K}^{(m)T}\mathbf{K}^{(m)}\mathbf{c}^{(s)}$ is the average sum power through subcarrier $N-N_J$ to $N-1$ and $\mathbf{K}^{(m)}$ is the corresponding selection matrix, and $\bar{\mathbf{K}}=\frac{1}{M}(\mathbf{I}+\sum_{m=1}^M\mathbf{K}^{(m)T}\mathbf{K}^{(m)})$.
It is easily seen that this constraint is convex in $\mathbf{c}^{(s)}$.

If the channel is flat-fading, i.e., the diagonal matrix $\mathbf{\Sigma}_H= \mathbf{I}$, the optimization problem can be formulated as
the following convex optimization problem\footnote{The reason to consider this case is to show that the power gain achieved is universal, so it could help with Line-of-Sight (LOS) scenario.}
\begin{equation}
\begin{aligned}
& \underset{\mathbf{c}^{(s)},d_{min}}{\text{maximize}}
& & d_{min} \\
& \text{s.t.}
& & 2\mathbf{c}^{(s)T}_0\mathbf{E}_{pq}\mathbf{c}^{(s)}-
\mathbf{c}^{(s)T}_0\mathbf{E}_{pq}\mathbf{c}^{(s)}_0\geq d^2_{min}, \forall (p,q)\\
&&& \boldsymbol{\phi}_n^T\mathbf{J}^{(m)}\mathbf{c}^{(s)}\geq 0, \qquad\forall (n,m)\\
&&& \mathbf{c}^{(s)T}\bar{\mathbf{K}}\mathbf{c}^{(s)}+M_fP_{Indep}\leq P.
\end{aligned}\label{17}
\end{equation}
The optimal solution can be obtained using standard solver such as CVX \cite{cvx} embedded with MATLAB.
A solution to the original optimization problem without approximation (\ref{equ.Approximation})
can be obtained via setting different initializations $\mathbf{c}^{(s)}_0$.

If the channel is selective-fading, we apply a block diagonal linear pre-equalizer $[\mathbf{F}]_{(2N_J+1)\times (2N_J+1)}=blkdiag\{f_0,\mathbf{F}_1,\ldots,\mathbf{F}_{N_J}\}$ before the IFFT block,
where the channel matrices $\mathbf{F}_k$ and the DC gain $f_0$ are  \cite[Prop.1]{Qian}
\begin{equation}
\mathbf{F}_{k\in[1,N_J]}=\frac{\alpha}{|z_{k}|^2}\cdot
\begin{bmatrix}
\Re(z_k) & -\Im(z_k)   \\
\Im(z_k) & \Re(z_k)
\end{bmatrix},~~f_0=\frac{\alpha}{z_0},
\end{equation}
and $z_k=\sum_i\beta_ie^{-j\frac{2k\pi\tau_i}{N}}$ with $\beta_i$ and $\tau_i$ denoting the gain and delay of $i$-th path respectively, $\alpha$ is a scaling factor that compensates for the path-loss component. In this case, the constellation needs to be designed via the following optimization,
\begin{equation}
\begin{aligned}
& \underset{\mathbf{c}^{(s)},d_{min}}{\text{maximize}}
& & d_{min} \\
& \text{s.t.}
& & 2\mathbf{c}^{(s)T}_0\mathbf{E}_{pq}\mathbf{c}^{(s)}-
\mathbf{c}^{(s)T}_0\mathbf{E}_{pq}\mathbf{c}^{(s)}_0\geq d^2_{min}, \forall (p,q)\\
&&& \boldsymbol{\phi}_n^T\mathbf{J}^{(m)}\mathbf{F}^{(s)}\mathbf{c}^{(s)}\geq 0, \qquad\forall (n,m)\\
&&& \mathbf{c}^{(s)T}\mathbf{F}^{(s)T}\bar{\mathbf{K}}\mathbf{F}^{(s)}\mathbf{c}^{(s)}+M_fP_{Indep}\leq P,
\end{aligned}\label{19}
\end{equation}
where $\mathbf{F}^{(s)} \triangleq \mathbf{F}\otimes \mathbf{I}_M$.

\section{Simulation Results}
We show an illustrative design example, where the key system parameters are chosen as follows: $N=8$, $N_J=3$ (full) thus $P_{Indep}=0$, $N_b=6$ and thus $M=64$.
For a regular DCO-OFDM system in comparison, we assume $M_f=3$, and that the subcarrier symbols take value from $[\pm 1\pm j]$ thus $P_{Indep}=2$ and MED equals $2$.
A fixed bias $B_{DC}$ can either be determined by exhaustive calculation according to all possible combinations of IFFT input symbols,
or by applying equation $B_{DC,z\text{dB}}=k\sqrt{\mathbb{E}\{x(t)^2\}}$.
Since $N$ is small, the complexity of exhaustive search is affordable and thus employed.
Note that, for large $N$, a 7dB or 13dB DC-bias is typically employed.
For the above-mentioned DC bias, the associated sum powers of DCO-OFDM are denoted as $P_{E}$, $P_{7dB}$ and $P_{13dB}$, respectively;
and the same average powers are employed for independent constellation design.

Heuristically, the DCIO-OFDM can outperform the DCO-OFDM, only when $d_{min}>2$ is achieved\footnote{Since $d_{min}^2$ equals the integration of amplitude difference square of two neighboring symbols over one interval, it has the same unit with $N_0$.},
since the SER under sufficient signal-to-noise ratio can be written as
\begin{equation}
E_s\approx \alpha Q\bigg(\sqrt{\frac{d_{min}^2}{2N_0}}\bigg),
\end{equation}
where $N_0$ is the power spectrum density of noise, $\alpha$ is a scaler depending on constellation size, the number of nearest constellation neighbors, etc.,
and $Q(\cdot)$ is the Gaussian Q-function which governs the SER.
The bit error rate associated is denoted by $E_{b}=\frac{\lambda}{N_b}E_{s}$,
where $\lambda$ is the number of bits in error for each symbol in error.
On average, around $2$ bits are off target for every wrong symbol carrying $6$ bits with the help of this mapper.
The details are shown in~\cite{Qian} and thus omitted here for conciseness.

We solve \eqref{17} via setting $100$ points of $\mathbf{c}^{(s)}_0$ for initialization,
for a flat-fading channel with $\boldsymbol{\Sigma_H}=\mathbf{I}$.
We consider the following three scenarios:
\begin{itemize}
  \item when the minimum bias $B_{DC,E}=2.14$ is applied for DCO-OFDM, the lowest power $P_{DCO,E}=16.58$ is associated, and the obtained MED for IDCO-OFDM when $P_{IDCO}=P_{DCO,E}$ is $d_{min}=2.27$;
  \item  when the medium (7dB) bias $B_{DC,7\text{dB}}=2.45$ is applied for DCO-OFDM, the medium power $P_{DCO,7\text{dB}}=18.01$ is associated, and the obtained MED for IDCO-OFDM when $P_{IDCO}=P_{DCO,7\text{dB}}$ is $d_{min}=2.37$;
  \item  when the high (13dB) bias $B_{DC,13\text{dB}}=5.33$ is applied for DCO-OFDM, the lowest power $P_{DCO,13\text{dB}}=40.41$ is associated and the obtained MED for IDCO-OFDM when $P_{IDCO}=P_{DCO,13\text{dB}}$ is $d_{min}=3.55$.
\end{itemize}
The bit-error-rate versus signal-to-noise ratios for the above threes scenarios are shown in Fig. 2.
It is observed that, at BER requirement $10^{-5}$, around $0.5$dB power gain is achieved by IDCO-OFDM with minimum power,
and this gain goes beyond $1$dB with medium and high powers.

\begin{figure}[ht]
\centerline{\includegraphics[width=1.1\columnwidth]{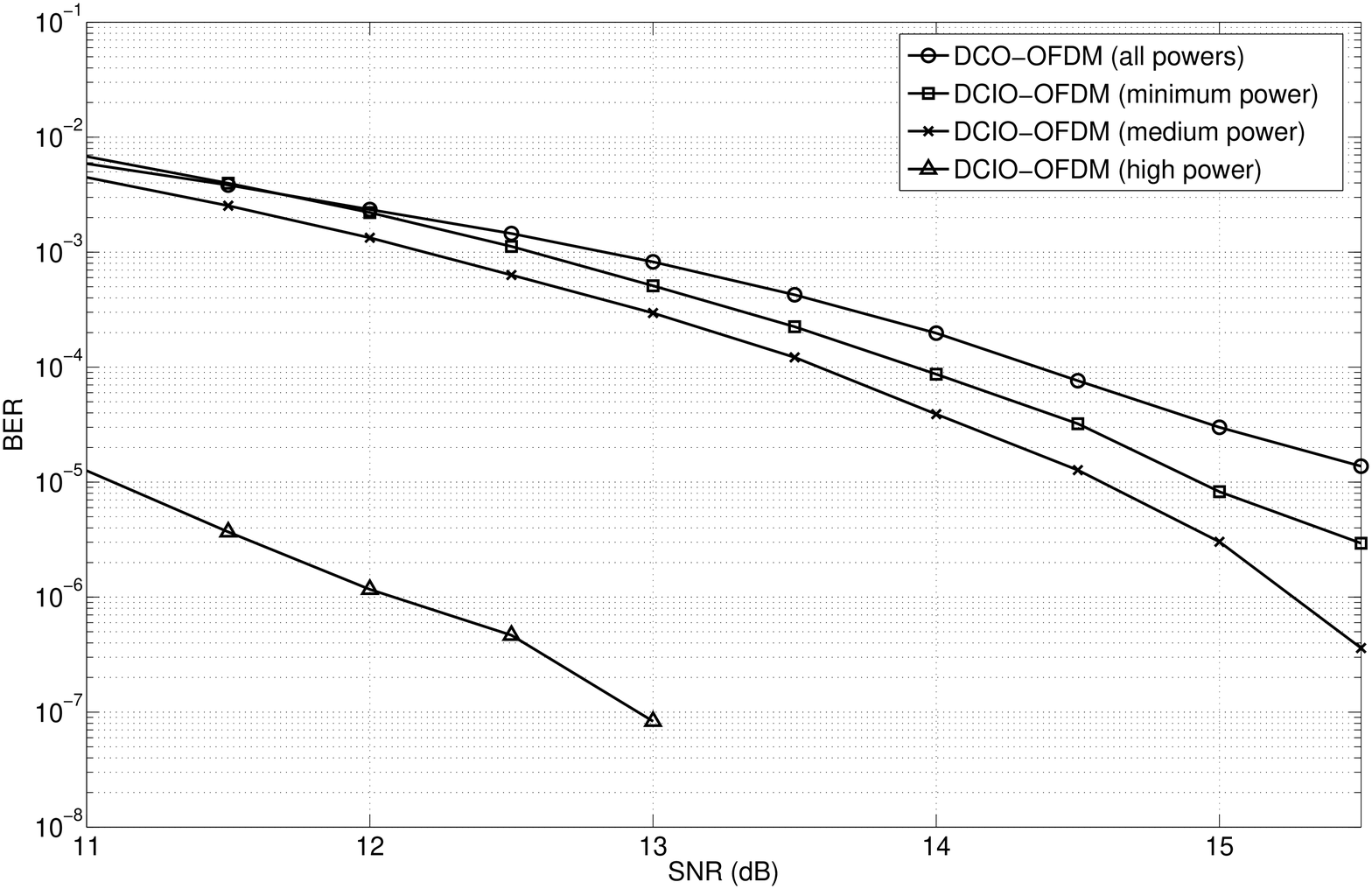}}
\vspace{-0.1in}
\caption{BER vs SNR for the DCO-OFDM and DCIO-OFDM systems with different powers.}
\end{figure}

For frequency-selective fading channels, similar gain of the DCIO-OFDM is observed.
The details are omitted here due to the lack of space.

\section{Conclusion and Future Works}
We have proposed a novel IFFT-based optical OFDM scheme which uses the DC-bias to carry information. In a high dimensional space including the DC and multiple subcarriers, optimized constellation is obtained through solving a convex optimization problem. Further, a BSA scheme is adopted to further reduce the system BER.
Non-negligible power gains are observed from simulations.
We believe that the idea of proposed DC-informative constellation design can be adopted
to other optical modulation schemes.

\end{document}